\documentclass[10pt,a4paper,twocolumn]{article}
\usepackage{fukugo}
\usepackage{amsmath}
\usepackage{amsfonts}
\usepackage{amssymb}
\begin{document}
%
\title{Neutrino masses in SU(5) grand unified model}
%
%
%

\author{{\large{Noriyuki Oshimo}} \medskip \\ 
  \it{ Department of Physics and Chemistry,} \\ 
  \it{ Graduate School of Humanities and Sciences,       } \\ 
  \it{ Ochanomizu University                             }
       }
%
%
%
\maketitle
\pagestyle{empty}
\thispagestyle{empty}
\setlength{\baselineskip}{13pt}
%
%

%
\def\barT{\overline T}
\def\barU{\overline U}
\def\barH{\overline H}
\def\barPsi{\overline\Psi}

     One of the most important issues in particle physics at present is 
neutrino masses and lepton generation mixing.   
Various experiments suggest that the neutrinos have extremely 
small but non-vanishing masses.  
Furthermore, the generation mixing for leptons is not suppressed, 
contrary to that for quarks.   
The Maki-Nakagawa-Sakata (MNS) matrix for the leptons is very 
different from the Cabibbo-Kobayashi-Maskawa (CKM) matrix for the quarks.  
The standard model (SM) must be necessarily extended.   

\smallskip

     Within the framework of grand unified theories (GUTs) coupled 
to supersymmetry, the small neutrino masses are usually explained  
by introducing right-handed neutrinos with large Majorana masses.  
This scenario is embodied in the GUT models whose gauge group   
is SO(10) or a larger one, since the minimal GUT group of SU(5) 
does not contain a representation for the right-handed neutrinos.     
However, the CKM matrix and the MNS matrix are generically related 
to each other in these models. 
Their observed difference is not trivially understood.   
Various approaches have been tried to explain the 
difference [NO1], though contrived schemes are often invoked.  

\smallskip

     For accommodating the neutrino masses and lepton generation mixing 
naturally, I propose a simple model [NO2] based on SU(5) 
and $N=1$ supergravity.   
Although right-handed neutrinos are not involved,  
left-handed neutrinos have small Majorana masses.   
The CKM matrix and the MNS matrix become independent of each other.  
Their observed difference is merely due to different values for free 
parameters.  
The ordinary SU(5) model encounters a difficulty
in describing masses of some down-type quarks or charged leptons.  
A solution of this problem is also provided.   

\smallskip

     The model consists, for Higgs superfields, of $\Phi$, $H$, $\barH$, 
$T$, $\barT$, $U$, and $\barU$, whose dimensions are respectively given by 
$\bf 24$, $\bf 5$, $\overline{\bf 5}$, $\bf 15$, $\overline{\bf 15}$, 
$\bf 45$, and $\overline{\bf 45}$.  
The matter superfields are represented by $\Psi_i$ of $\bf 10$ and 
$\barPsi_i$ of $\overline{\bf 5}$, with $i$ being the generation index.  
In addition to the particle contents of the minimal SU(5) model, 
superfields of $\bf 15$, $\overline{\bf 15}$, $\bf 45$, and 
$\overline{\bf 45}$ are introduced.  
This $\bf 15$ representation couples to 
$\overline{\bf 5}\times\overline{\bf 5}$ of matters, leading to Majorana 
masses for left-handed neutrinos.  
The $\overline{\bf 45}$ representation gives masses to down-type quarks 
and charged leptons through its coupling to the matters 
$\bf 10\times\overline{5}$.  
The complex conjugate representations $\overline{\bf 15}$ and $\bf 45$  
are also incorporated for having mass terms and anomaly cancellation.  

\smallskip

     The superpotential is the sum of all 
renormalizable terms consistent with SU(5) and $R$ parity, 
$W=W_M+W_H$.  
The couplings of matter superfields to Higgs superfields are contained 
in $W_M$, which is given by 
$$
\begin{aligned}
   W_M &= \Gamma_{ij}^u\epsilon H\Psi_i\Psi_j  
         + \Gamma_{ij}^{de}{\barH}\Psi_i{\barPsi_j}   
         + \Gamma_{ij}^\nu {\barPsi_i}T{\barPsi_j}  \\ 
  &+  \overline{\Gamma}_{ij}^{u}\epsilon U\Psi_i\Psi_j +  
      \overline{\Gamma}_{ij}^{de}\barU\Psi_i\barPsi_j, 
\end{aligned}
$$
with $\epsilon$ being the totally antisymmetric tensor of rank five.  
If the Higgs superfield $\barU$ is not involved, the down-type quark and 
the charged lepton would have the same mass in each generation.  
The superpotential $W_H$ contains only Higgs superfields.  
There are three degenerate vacua under global supersymmetry, 
though this degeneracy is lifted by $N=1$ supergravity.  
The gauge symmetry of the GUT is broken down to that of the SM 
by the vacuum expectation value (VEV) 
$\langle \Phi \rangle= {\rm diag}(1, 1, 1, -3/2,-3/2)v_\Phi$.  
The masses of $X$ and $Y$ bosons are given by 
$M_X=M_Y=(5/2\sqrt{2})g_5v_\Phi$, which are typically of the 
order of $10^{16}$ GeV.    

\smallskip

     Below the GUT energy scale, the superpotential for the quark 
and lepton masses within the framework of SU(3)$\times$SU(2)$\times$U(1) 
is written as 
$$
\begin{aligned}
W_M &= \eta_d^{ij} H_1\epsilon Q^iD^{cj} + \eta_u^{ij}H_2\epsilon Q^iU^{cj}
           \\
 &+ \eta_e^{ij} H_1\epsilon L^iE^{cj} 
 +  \frac{1}{2}\kappa^{ij}(\epsilon L^i)^T T\epsilon L^j,  
\end{aligned}
$$
where $\epsilon$ stands for the totally antisymmetric tensor of rank two.    
The superfields for the quarks and leptons are expressed by a 
self-explanatory notation.   
The Higgs superfields $H_1$ and $H_2$ are composed of the SU(2)-doublet 
components of $\bf 5$, 
$\overline{\bf 5}$, $\bf 45$, and $\overline{\bf 45}$.   
We assume that the masses of $H_1$ and $H_2$ become of the 
order of the electroweak energy scale by fine-tuning or 
some other mechanism.  
The remaining superfields in $H$, $\barH$, $U$, and $\barU$ have masses 
of the order of $M_X$.   
The SU(2)-triplet component of $\bf 15$ is denoted by the same symbol.   
Owing to a large magnitude of $\eta_u^{ij}$ corresponding to the $t$-quark 
mass, the gauge symmetry SU(2)$\times$U(1) is broken down to 
U(1)$_{\rm EM}$ through quantum corrections.
The Higgs bosons have the VEVs  
$\langle H_1\rangle =(v_1/\sqrt{2}, 0)$ and
$\langle H_2\rangle =(0, v_2/\sqrt{2})$,
where $v_1$ and $v_2$ are related to the $W$-boson mass as
$M_W=(1/2)g_2\sqrt{v_1^2+v_2^2}$.

\smallskip

     The left-handed neutrinos receive tiny Majorana masses.     
A non-vanishing VEV for $T$ is induced by the electroweak symmetry 
breaking, 
$$
\langle T \rangle={\rm diag}(0,v_T/\sqrt{2}), \quad  
 v_T = \frac{\lambda v_2^2}{\sqrt{2}m_T}, 
$$
where $\lambda$ and $m_T$ denote a dimensionless coupling constant 
and a mass parameter for $T$, respectively, in the Higgs potential 
derived from $W_H$.   
Assuming $\lambda\sim 1$, the value of $v_T$ is of the order of 
$10^{-1}$~eV for $m_T\sim 10^{14}$ GeV.  
The neutrinos can have the masses which are of the correct order 
of magnitude.   
The appropriate value for $m_T$ is smaller than $M_X$ by one or two 
orders of magnitude.  

\smallskip

     As the physical parameters for the coefficients $\eta_d$, $\eta_u$, 
$\eta_e$, and $\kappa$, we can take the CKM matrix $V_{CKM}$, 
the MNS matrix $V_{MNS}$, and the diagonalized eigenvalue matrices 
$\eta_d^D$, $\eta_u^D$, $\eta_e^D$, and $\kappa^D$.   
These matrices depend on the energy scale of physics, 
which are governed by the renormalization group equations.  
If the matrices at the electroweak energy scale are measured experimentally, 
those at the GUT energy scale are determined.  

\smallskip

     The superpotential $W$ at the GUT energy scale can accommodate 
any values for the CKM and MNS matrices.  
Any masses for the quarks and leptons can also be realized.  
For expressing the coefficient matrices, we take 
the generation basis in which $\eta_d$ and $\eta_e$ are diagonal.   
The required values for $V_{CKM}$, $V_{MNS}$, $\eta_d^D$, $\eta_u^D$, 
$\eta_e^D$, and $\kappa^D$ at the GUT energy scale are fulfilled 
by taking the coefficients for   
$$
\begin{aligned}
\Gamma^\nu &= \frac{1}{2}(V_{MNS})^T \kappa^D V_{MNS},  \\
    \Gamma^{de} &= -\frac{1}{2\sqrt{2}(C_1)_{11}}
                           \left(3\eta_d^D+\eta_e^D\right),  \\
    \overline{\Gamma}^{de} &= -\frac{\sqrt{3}}{2(C_1)_{21}}
                           \left(\eta_d^D-\eta_e^D\right),  \\
    \Gamma^u &= \frac{1}{8(C_2)_{11}}
       \Bigl\{(V_{CKM})^T\eta_u^D  + \eta_u^DV_{CKM}\Bigr\},  \\
    \overline{\Gamma}^u &= -\frac{\sqrt{6}}{8(C_2)_{21}}
       \Bigl\{(V_{CKM})^T\eta_u^D - \eta_u^DV_{CKM}\Bigr\},  
\end{aligned}
$$
where $C_1$ and $C_2$ denote the unitary matrices which prescribe 
the structures of $H_1$ and $H_2$.  
These equations can always be satisfied, since the coefficients 
$\Gamma^\nu$, $\Gamma^{de}$, $\overline{\Gamma}^{de}$, 
$\Gamma^u$, and $\overline{\Gamma}^u$ are independent mutually.   
The MNS matrix is not constrained by the CKM matrix.  

\smallskip

     In the minimal SU(5) model, the gauge coupling unification infers 
too rapid proton decay.  
However, in the present model, there are additional Higgs superfields 
which affect the evolutions of the gauge coupling constants  
around the GUT energy scale.   
The relation between the decay and the unification becomes relaxed, 
and the proton life time could be long sufficiently.   

\smallskip

     This work is supported in part by the Grant-in-Aid for
Scientific Research on Priority Areas (No. 14039204) from the
Ministry of Education, Science and Culture, Japan.

\smallskip

\end{document}